\documentclass{article} 
\usepackage{epsfig}

\begin{document}

\title{Wigner-function description of quantum teleportation in
  arbitrary dimensions and continuous limit} 
\author{M. Koniorczyk$^{1,2}$, V. Bu\v zek $^{3,4}$ and J. Janszky$^1$\\
$^1$Department of Nonlinear and Quantum Optics,\\
    Research Institute for Solid State Physics and Optics,\\
    Hungarian Academy of Sciences,\\
    P.O. Box 49, H-1525 Budapest, Hungary
\\$^2$Institute of Physics, 
    University of P\'ecs, \\
    Ifj\'us\'ag \'ut 6. H-7624 P\'ecs, Hungary 
\\$^3$Institute of Physics,\\
    Slovak Academy of Sciences
    D\'ubravsk\'a cesta 9,\\
    842 28 Bratislava, Slovakia
\\$^4$Faculty of Informatics,\\ Masaryk University,\\
Botancik\' a 68 a, Brno 602 00, Czech Republic}
\date{3 April 2001}

\maketitle

\begin{abstract}
We present a unified approach to quantum teleportation in
  arbitrary dimensions based on the Wigner-function formalism. This
  approach provides us with a clear picture of all
  manipulations performed in the teleportation protocol. In addition
  within the framework of the Wigner-function formalism all the
  imperfections of the manipulations can be easily taken into
  account.
\end{abstract}

%\draft

%\begin{multicols}{2}

All quantum mechanical phenomena may be described in terms of
quasiprobability distributions, as an alternative to the direct
application of density matrices. Wigner functions are especially
frequently applied, as they behave similarly to classical probability
distributions from several points of view. For quantum states with
infinite dimensional Hilbert-spaces, the application of
Wigner functions has become a standard part of considerations.  For
finite dimensional Hilbert-spaces, the Wigner-function formalism was
first investigated by Wootters \cite{annph176_1}. The discrete
Wigner functions have shown to be useful in investigating coherent
states in a finite-dimensional basis \cite{pra45_8079}, definition of
Q-functions and other propensities \cite{pra52_2419}, and also played
role in the development of number-phase Wigner functions
\cite{pra41_5156}. Quantum tomography for finite-dimensional
Wigner functions has also been developed, applying a generalized
definition \cite{prl74_4101}.

A great deal of attention has been paid recently to the phenomenon of
quantum teleportation, which is the basic primitive of quantum
communication, and it is also interesting from the point of view of
quantum nonlocality \cite{pra62_032101}.  The experimental feasibility
of the phenomenon
\cite{nature390_575,prl80_1121,science282_706,nature396_52} highly
contributes to the importance of these investigations. The idea of
quantum teleportation by Bennett \cite{prl70_1895}, was formulated on
finite dimensional Hilbert-spaces. In this context, the conventional
description applying Hilbert-space vectors is appropriate. On the
other hand, the idea of continuous variable quantum teleportation,
proposed originally by Vaidman \cite{pra49_1473}, was first put into a
quantum optical context by Braunstein and Kimble using the
Wigner-function formalism \cite{prl80_869}.  However, this scheme may
also be described in terms of either wavefunctions
\cite{pra60_937,pra61_032302} or Fock-states\cite{pra60_5095}, and a
low-dimensional coherent state description has also been developed
recently \cite{conttelep}. A covariant description in terms of
canonically conjugate observables and their eigenstates is also
possible \cite{pra61_022310}, providing a description valid for both
discrete and continuous dimensions.

In this paper we present the description of quantum teleportation
purely in the framework of Wigner-function formalism of quantum
mechanics. The main emphasis is put on the case of finite dimensional
Hilbert-spaces, but we make some comments on the infinite dimensional
limits.  It will be shown, that the entire process of quantum
teleportation can be consistently described purely in terms of
Wigner functions, and in this context, the finite and infinite
dimensional cases can be treated in a conceptually uniform way.

The paper is organized as follows: After a brief review of some
elements of finite dimensional Wigner-function formalism, we describe
the ideal EPR state. Then the entire teleportation process is
discussed, and conclusions are drawn.

Consider a physical system, with states described by the
$N$-dimensional Hilbert-space ${\cal H}$. We define two non-commuting
Hermitian operators $\hat q$ and $\hat p$ describing two canonically
conjugate quantities. We will call them ``position'' and ``momentum''
respectively, though they may be realized by several physical
quantities, as, for instance, photon number and Pegg-Barnett phase
operators on a truncated Fock-space. The operators are defined as:
\begin{equation}
  \label{eq:q}
  \hat q=\sum_{k=0}^{N-1}k|k\rangle\langle k|, 
  \qquad 
  \hat p=\sum_{l=0}^{N-1}l|p_l\rangle\langle p_l|
\end{equation}
where the set of $|k\rangle$ position and $|p_l\rangle$ momentum
eigenstates both form an orthonormal basis on ${\cal H}$, and
\begin{equation}
  \label{eq:pseig}
  |p_l\rangle =\frac{1}{\sqrt{N}} \sum_{k=0}^{N-1} e^{i\frac{2\pi
   }{N}kl}
|k\rangle
\end{equation}
holds. 

Wigner functions for this discrete system can be defined in a slightly
different manner depending on the properties of the number $N$, the
dimensionality of the corresponding Hilbert space. In what follows we
will suppose, that $N$ is greater than 3 and it is a prime number.
Though it introduces some loss of generality, apart from technical
details, there is no significant physical difference between the cases
discussed, and the remaining two possibilities. In case of $N=2$, a
different definition of the Wigner function has to be applied, while
for composite $N$-s, the phase spaces are Cartesian products of lower
dimensional phase-spaces. Alternatively, one may use the formalism
suggested in Ref.~\cite{prl74_4101}.

According to the original paper of Wootters \cite{annph176_1}, the
Wigner function corresponding to a state in a Hilbert space with
dimension $N\geq 3$ prime, is defined with the aid of the discrete
Wigner operator
\begin{equation}
  \label{eq:A}
  \hat A(q,p)=\sum_{r,s}
\delta_{2q,r+s}e^{i\frac{2\pi}{N}p(r-s)}|r\rangle\langle s|,
\end{equation}
where $q$ and $p$ take integer values from $0$ to $N-1$. The $(q,p)$
pairs constitute the discrete phase space.  For a state described by a
density matrix $\varrho$ the Wigner function is
\begin{equation}
  \label{eq:W}
  W(q,p)=\frac{1}{N}\mathop{\mbox{Tr}} (\varrho \hat A).
\end{equation}
Wigner functions defined in this way obey analogous properties to
those defined on infinite dimensional Hilbert-spaces.  The marginal
distributions of the functions
\begin{equation}
  \label{eq:marg}
  P_q(q)=\sum_{p}W(q,p) ,  \qquad P_p(p)=\sum_{q}W(q,p)
\end{equation}
describe the statistics of measurements of observables $\hat q$ and
$\hat p$ respectively.  

For multipartite systems, Wigner functions are defined, similarly to
the infinite dimensional case, with the expectation values of the
direct product of the Wigner operators. In what follows we consider
multipartite systems with Hilbert-spaces of equal dimension. For a
bipartite system with subsystems 1 and 2, described by the joint
density matrix $\varrho^{(12)}$,
\begin{equation}
  \label{eq:2W}
  W(q_1,p_1,q_2,p_2)=
\frac{1}{N^2} \mathop{\mbox{Tr}} (\varrho^{(12)}
\hat A_1(q_1,p_1) \otimes \hat A_2(q_2,p_2))
\end{equation}
Wigner functions describing a subsystem are obtained by summing the
joint Wigner function in the corresponding set of the respective
variables, e. g. from Eq.~(\ref{eq:2W}) we have
\begin{eqnarray}
  \label{eq:parctr}
W(q_1,p_1)=\sum_{q_2,p_2=0}^{N-1}W(q_1,p_1,q_2,p_2), \nonumber \\
W(q_2,p_2)=\sum_{q_1,p_1=0}^{N-1}W(q_1,p_1,q_2,p_2).
\end{eqnarray}

For bipartite systems, the completely entangled Bell-states
\begin{equation}
  \label{eq:bell}
  |\Xi_{p,x}\rangle_{12}=\frac{1}{\sqrt{N}}\sum\limits_{k=0}^{N-1}
e^{i\frac{2\pi}{N}kp}|k\rangle_1|k-x\rangle_2,
\end{equation}
form an orthonormal basis on the ${\cal{H}}\otimes{\cal{H}}$
Hilbert-space of the joint system. These are common eigenstates of the
following joint observables:
\begin{eqnarray}
  \label{eq:belleig}
  (\hat q_1-\hat q_2)|\Xi_{p,x}\rangle_{12}=(q_1-q_2)|\Xi_{p,x}\rangle_{12},
  \nonumber \\
  (\hat p_1+\hat p_2)|\Xi_{p,x}\rangle_{12}=(p_1+p_2)|\Xi_{p,x}\rangle_{12}.
\end{eqnarray}

Following Bennett \cite{prl70_1895}, we shall suppose that the sender,
Alice, and the receiver, Bob, share the subsystems 2 and 3 in the
entangled state
  \begin{equation}
    \label{eq:bellvect}
    |\Xi_{0,0}\rangle_{23}=\frac{1}{\sqrt{N}}
\sum\limits_{k=0}^{N-1}|k\rangle_2|k\rangle_3.
  \end{equation}
In what follows, we shall use the term ``EPR-state'' for this state.
The Wigner function of this state 
can be calculated according to Eqs.~(\ref{eq:A}),
(\ref{eq:W}) and (\ref{eq:2W}), and is found to be
\begin{equation}
  \label{eq:WEPR}
  W_{\mathrm{EPR}}(q_2,p_2,q_3,p_3)=
  \frac{1}{N^2}\delta_{q_2,q_3}\delta_{p_2,-p_3}.
\end{equation}

Calculating the Wigner functions for subsystems 2 and 3 according to
Eq.~(\ref{eq:parctr}), both of them are found to be the constant
$1/N^2$.  From this follows, that any of the marginals describe a
uniform distribution. This reflects the EPR nature of the state:
making observations on either of the subsystems separately, both
position and momentum have random values. On the other hand, according
to Eq.~(\ref{eq:belleig}), some joint observables have definite
value, as it is also clearly reflected by Eq.~(\ref{eq:WEPR}):
$q_2-q_3=0$ and $p_2+p_3=0$. From this we may conclude, that the form
of EPR Wigner function in Eq.~(\ref{eq:WEPR}) could have been even a
plausible ansatz.

The Wigner function in Eq.~(\ref{eq:WEPR}) shows the connection with
the EPR state used by Braunstein and Kimble for continuous variable
teleportation. In the continuous variable case, for an ideal EPR state
Dirac-deltas appear, corresponding to a state with infinite energy.
Therefore instead of the ideal EPR state, usually two-mode squeezed
vacuum is considered instead, which results in the imperfection of the
protocol.

Let us consider the teleportation process. Alice, the sender and Bob
the receiver have shared the EPR pair described by the Wigner function
in Eq.~(\ref{eq:WEPR}). In addition 
Alice has system $1$ in the arbitrary state
described by a Wigner function $W_{\mathrm{in}}(q_1,p_1)$. 
The joint Wigner function of the whole system is thus
\begin{equation}
  \label{eq:Wbe}
  W(q_1,p_1,q_2,p_2,q_3,p_3)=
  \frac{1}{N^2}W_{\mathrm{in}}(q_1,p_1)\delta_{q_2,q_3}\delta_{p_2,-p_3}
\end{equation}
Alice has to carry out a projective measurement on subsystems 1 and 2.
This measurement is performed in the Bell basis which obviously projects
the systems 1 and 2 on the Bell states (\ref{eq:bell}). As we have
already mentioned these states 
are simultaneous eigenstates of the joint observables $\hat
X_2=\hat q_1- \hat q_2$ and $\hat P_1=\hat p_1+ \hat p_2$. In order to
describe the measurement, we have to express the Wigner function in
Eq.~(\ref{eq:Wbe}) in terms of these variables and $X_1=\hat q_1+ \hat
q_2$ and $\hat P_2=\hat p_1- \hat p_2$, instead of $q_1,p_1$ and
$q_2,p_2$. Note, that because of the modulo $N$ arithmetics, the
ranges of the new variables are the same.

This canonical transformation is more straightforward in the infinite
dimensional case, where we can introduce a $\frac{\sqrt{2}}{2}$ factor
in the definition of the new variables, and thus it is easy to express
the inverse transformation in the same fashion. In our case, a division
by $2$ appears in the inverse formula, which seems to be inappropriate
at first sight. This problem can be overcome in the following way: As
$N$ is odd, we may introduce a ``generalized division by $2$'' in the
modulo $N$ sense as
\begin{equation}
  \label{eq:div2}
  {\cal {D}}_2(k)=\cases{ 
\frac{k}{2} ,\quad n\ \mathrm{even} \cr 
\ \cr
\frac{k+N}{2},\quad n\ \mathrm{odd}},
\end{equation}
which has the property $2{\cal {D}}_2(k)=k$.
Here we emphasize again, that {\it all} additions, subtractions and
multiplications are understood in the modulo $N$ sense.
With the aid of this operation, the old variables can be expressed as
\begin{eqnarray}
  \label{eq:traf}
  q_1={\cal {D}}_2(X_1+X_2),\qquad q_2={\cal {D}}_2(X_1-X_2)\nonumber \\
  p_1={\cal {D}}_2(P_1+P_2),\qquad p_2={\cal {D}}_2(P_1-P_2).
\end{eqnarray}
The Wigner function in Eq.~(\ref{eq:Wbe}) after the transformation is
\begin{eqnarray}
  \label{eq:JWig}
& &  W(X_1,P_1,X_2,P_2,q_3,p_3)=
  \frac{1}{N^2}
  \delta_{X_1-X_2,2q_3}\delta_{P_1-P_2,-2p_3}
\nonumber \\
&\times &
W_{\mathrm{in}}({\cal {D}}_2(X_1+X_2),{\cal {D}}_2(P_1+P_2)).
\end{eqnarray}
At this stage, all subsystems are entangled. Note, that the canonical
transformation, which is described here by introducing new variables,
is physically a unitary transformation which entangles two
subsystems, and it even cannot be carried out completely by using
passive linear optical elements \cite{pra59_3295}, and may require nonlinear
optics \cite{prl86_1370}.

Now we are ready to describe the Bell-state measurement, which results
in values $X_2$ and $P_1$, the classical information, which is sent to
Bob. Summing the Wigner function in Eq.~(\ref{eq:JWig}) in variables
$X_1,P_2,q_3,p_3$, we obtain the probability distribution of the
measurement results, which is equal to constant $1/N^2$. Thus we can
obtain each possible measurement result with equal probability, in
accordance with Bennett's description. 

\begin{figure}
\begin{center} 
\epsfig{file=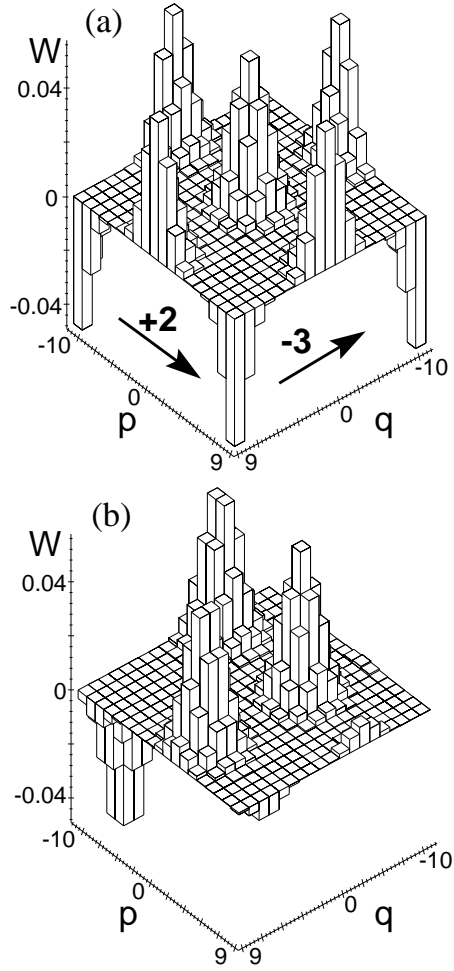,width=6cm}
\bigskip
  \caption{
    Shifting of Wigner function in a discrete phase space of a quantum system
with a $19$-dimensional Hilbert space.
    (a) shows the state, which is a discrete counterpart of the
    harmonic
    oscillator ground-state (see Ref.~ \protect\cite{pra52_2419}). (b)
    is shifted version, according to the arrows in figure (a). Points
    of the phase space are indexed so that the main peak is centered in
    the origin of a phase space; recall the modulo $N$ summation.}
  \label{fig:figure}
\end{center}
\end{figure}

To describe the conditional projection by the measurement, we have to
keep variables $X_2$ and $P_1$, constants, as these numbers constitute
the result of the measurement, and we have to sum the Wigner function
of Eq.~(\ref{eq:JWig}) in variables $X_1$ and $P_2$, as we loose all
information about these because of the projective measurement. This
procedure is the exact analogue of the continuous case. The resulting
Wigner function has to be renormalized, and it has the form
\begin{equation}
  W_{\mathrm{out}}(q_3,p_3)=W_{\mathrm{in}}(q_3+X_2,p_3+P_1).
\end{equation}
It is seen, that the resulting Wigner function is a shifted
version of the original, and the shift is determined by the result of
the measurement. This is the exact analog of the continuous case. 
Bob, possessing the values $X_2$ and $P_1$, can restore the the
teleported state. The shift in a finite dimensional Hilbert-space
is illustrated in Fig.~\ref{fig:figure}. Obviosuly, these shifts
correspond to translations (canonical transformations)
in a discrete phase space.

The required inverse transformation as described by Bennett is
\begin{equation}
  \label{eq:UBennett}
  U_{X_2,P_1}=\sum_k e^{i\frac{2\pi}{N}P_1k}|k\rangle \langle k-X_2|.
\end{equation}
It is easy to verify, that this transformation acts on a
Wigner function as
\begin{equation}
  \label{eq:Btraf_wig}
  W'(q,p)=\langle U^\dag_{X_2,P_1}A(q,p)U_{X_2,P_1}\rangle=W(q-X_2,p-P_1),
\end{equation}
thus our description is perfectly consistent with Bennett's results.

The similarity of our discussion to the original description of
continuous variable quantum teleportation by Braunstein and Kimble is
apparent. Care should be taken however, if the actual infinite
dimensional limit is be constructed from the description above, which
is far from straightforward indeed.  For instance, several nontrivial
problems have to be overcome if $\hat q$ and $\hat p$ is associated
with photon numbers and Pegg-Barnett phase
\cite{physscr_t48_94,pra52_3474}.

In conclusion, we have shown, that quantum teleportation can be
described purely in terms of Wigner functions, and this could have
been possible even without mentioning the underlying Hilbert-space.
This approach has several advantages in the description of
imperfections.  Noisy entanglement can be treated, similarly to the
continuous case, by replacing the Kronecker-deltas describing ideal
entangled states with the appropriate Wigner function. While
projective measurement is described by filtering with delta-functions
here, a fuzzy measurement may be described by filtering with unsharp
filters.  This example suggests, that Wigner functions may prove to be
a useful tool for investigating phenomena in multipartite systems with
finite dimensional Hilbert-spaces.

\section*{Acknowledgements} 

This work was supported by the IST projects EQUIP (IST-1999-11053) and
QUBITS (IST-1999-13021), and by the Research Fund of Hungary (OTKA)
under contract No. T034484.

%\bibliographystyle{prsty}
%\bibliography{optics}

%\end{multicols}

\end{document}